\documentclass[twocolumn,prb,floatfix]{revtex4-1}
\usepackage{amssymb,amsmath,latexsym}
\usepackage{bm}
\usepackage[pdftex]{graphicx}

\begin{document}

\title{Curvature-induced spin-orbit coupling and spin relaxation in a chemically clean single-layer graphene}
\author{Jae-Seung Jeong}
\email{jsjeong@kias.re.kr}
\affiliation{School of Physics, Korea Institute for Advanced Study, Seoul 130-722, Korea}
\author{Jeongkyu Shin and Hyun-Woo Lee}
\affiliation{Department of Physics and PCTP, Pohang University of Science and Technology, Pohang 790-784, Korea}

\date{\today}


\begin{abstract}

The study of spin-related phenomena in materials
requires 
knowledge of the 
precise form of effective spin-orbit coupling for conducting carriers in solid-states systems.
We demonstrate theoretically that 
curvature induced by corrugations or periodic ripples in single-layer 
graphenes generates two types of effective spin-orbit couplings.
In addition to the spin-orbit coupling reported previously
that couples with sublattice pseudospin and corresponds to the Rashba-type spin-orbit coupling
in a corrugated single-layer graphene,  
%
there is an additional spin-orbit coupling that does not couple with the pseudospin, 
which 
can not be obtained from the extension of the curvature-induced spin-orbit coupling of carbon nanotubes. 
Via numerical calculation we show
that both types of the curvature-induced spin-orbit coupling
make the same order of contribution to 
spin relaxation  
in chemically clean single-layer graphene with
nanoscale corrugation.
The spin relaxation dependence on the corrugation roughness is also studied. 
\end{abstract}


\maketitle

\section{Introduction}
Graphene has attracted much interest
due to its unusual linear energy spectrum and electronic properties.~\cite{RMP09Neto}
%
%
It is also a promising material
for spin-based applications
since spin relaxation (SR) in graphenes 
is expected 
to be weak
due to the suppression of the hyperfine interaction of $^{12}$C
and 
the weak atomic spin-orbit coupling (SOC). 
Various spintronic devices~\cite{Nature06Son} and spin qubits~\cite{NaturePhys07Trauzettel} 
on graphenes
have been suggested.
%

Recent experiments, however, 
measured SR times (SRTs) 
of the order of
$0.1\!-\!0.5\,$ns,~\cite{Nature07Tombros}
which were much shorter than expected
and comparable to other materials such as aluminum~\cite{Nature02Jedema}
and copper.~\cite{PRB03Jedema}
Those experiments stirred up intensive theoretical investigation
of the SR in single-layer graphene (SG)
to clarify the origin of the short SRT.
SG samples in Ref.~\cite{Nature07Tombros}
have low mobility $\sim\!2000\,{\rm cm^2/Vs}$
and the SRT was found to increase with increasing carrier density, 
which is consistent with 
the Elliot-Yafet (EY) SR mechanism~\cite{RMP04Zutic} arising from the
interplay of SOC and momentum scattering. 
However, it is not clear what kind of SOC is mainly
responsible for the SR in SG, 
and various kinds of SOCs are examined. 
It was suggested~\cite{PRL09Neto}
that adatoms on SG can enhance the local SOC 
to the order of $10\,$meV,
which results in short SRT comparable to the experimental value~\cite{Nature07Tombros}
for low mobility SG samples.
Thus the adatom-induced SOC provides 
a possible origin~\cite{PRB09Ertler,arXiv11Ochoa}
of the short SRT in Ref.~\cite{Nature07Tombros}.
If this SOC is indeed the main origin, 
which requires further confirmation, it implies that
the SRT may be enhanced in SGs without adatoms.
To find out how greatly the SRT can be enhanced,
various factors limiting the SRT should be examined carefully. 
In exfoliated SGs,  
charged impurities or surface phonons
in the substrate~\cite{PRB09Ertler} 
can induce the effective SOC of the order of $0.01\,$meV.
Also, 
effective SOC of the order of $0.01-0.1\,$meV~\cite{PRB06Huertas-Hernando}
can be induced by 
local curvature effects arising from 
corrugations~\cite{Nature07Meyer,PNAS07Stolyarova,NanoLett07Ishigami,PRL09Geringer,PRL10Cullen,PRB11Knox} and
periodic ripples~\cite{NatNano09Bao,Science11Choi} 
observed in suspended~\cite{Nature07Meyer,NatNano09Bao,PRB11Knox} 
and exfoliated~\cite{PNAS07Stolyarova,NanoLett07Ishigami,PRL09Geringer,PRL10Cullen,Science11Choi} SGs.
Each substrate-induced and curvature-induced SOC combined with
momentum scattering off impurities~\cite{arXiv11Ochoa,PRL09Huertas-Hernando}
or electron-electron Coulomb scattering,~\cite{PRB10Zhou}
and substrate-induced or curvature-induced SOC itself~\cite{PRB11Dugaev}
can cause SR. 
The estimated SRTs are at least of the order of $10\,$ns, 
which is about two order of magnitude larger than
measured SRTs.~\cite{Nature07Tombros}


For a more reliable investigation of
spin-related phenomena 
such as SR and spin coherence, 
it is important 
to know
the precise form of SOC.
%
In this respect, existing studies 
of the curvature-induced SOC in SGs
are not satisfactory, since the precise
form of this SOC is not known. 
Previous theoretical studies~\cite{PRB06Huertas-Hernando,PRL09Huertas-Hernando,PRB10Zhou,PRB11Dugaev}
assumed a particular form of the SOC
inferred
from that of a carbon nanotube (CNT).~\cite{PRB06Huertas-Hernando}
However the SOC-induced energy splitting measurement~\cite{Nature08Kuemmeth}
in ultra-clean CNTs revealed 
that the form of the SOC in Ref.~\cite{PRB06Huertas-Hernando,JPSJ00Ando}
does not provide a satisfactory description of the SOC effect in a CNT. 
Later theoretical works~\cite{JPSJ09Izumida,PRB09Jeong}
reported that the correct form of the SOC in a CNT 
has an additional term in addition to the SOC form in Ref.~\cite{PRB06Huertas-Hernando,JPSJ00Ando}.
The additional term does not couple to the pseudospin degrees of freedom 
in a CNT and thus differs qualitatively from the previously
known SOC term,~\cite{PRB06Huertas-Hernando,JPSJ00Ando}
which couples to the pseudospin. 
It was demonstrated in Ref.~\cite{PRB09Jeong}
that the interplay of the two SOC terms
can explain the SOC-induced energy splitting measurement.~\cite{Nature08Kuemmeth,NatPhys11Jespersen}
It is then reasonable to expect that 
a similar additional SOC term may exist for SGs as well.

Finding the correct curvature-induced SOC can have an implication not only on 
SR in the current experiments,~\cite{Nature07Tombros}
but also on SR in chemically clean SGs.  
As technology for the preparation of SGs progresses so as to 
diminish impurity effects, 
the fundamental limit of SR would be governed by intrinsic sources.
Recalling that atomically flat SGs are 
thermodynamically unstable,~\cite{NatMat07Fasolino}
the curvature-induced SOC is expected to be one of
the intrinsic origins determining the upperbound of the 
SRT in chemically clean SGs.~\cite{flatSG}
We believe 
that this expectation 
is very plausible since SGs and CNTs
are very similar chemically
and the dominant source
of the SOC in ultra-clean CNTs~\cite{Nature08Kuemmeth,NatPhys11Jespersen}
is the curvature-induced SOC. 






In this paper,
we show that in addition to the curvature-induced 
Rashba-type SOC reported 
in an existing theory,~\cite{PRB06Huertas-Hernando}
which couples to the pseudospin, 
there exists an additional type of
the curvature-induced SOC,
which does not couple to 
the pseudospin and thus
appears as the diagonal term in the pseudospin representation.
In this respect, this additional 
SOC term is similar to the additional SOC term 
in CNTs.~\cite{JPSJ09Izumida,PRB09Jeong}
However these two additional terms have
topologically different origins since the periodicity 
in the circumferential 
direction arising from the tube topology of CNTs
is absent in SGs. 
Thus the additional SOC term in SGs
cannot be obtained from 
the additional SOC term in CNTs
through a trivial extension. 
This point was not addressed in an earlier 
attempt~\cite{PRB09Jeong}
to derive the curvature-induced SOC in SGs. 
In this paper, we 
also examine the SR due to the curvature in chemically clean SGs
and show that the effect of the additional 
SOC on the SR is comparable 
in magnitude to the effect of the previously
known curvature-induced SOC. 
Thus the additional curvature-induced 
SOC term should be included
in studies of spin-related phenomena in SGs.
We also investigate SR dependence on 
the fractal dimension 
of the corrugation roughness,~\cite{NanoLett07Ishigami,PTTSA08Katsnelson}
and show that corrugation roughness affects the SR.

This paper is organized as follows. 
In Sec.~\ref{sec:CISOC},
we show analytical expressions of two types of the curvature-induced 
SOCs in SGs, and then demonstrate how to obtain them microscopically
using tight-binding (TB) Hamiltonian of the local curvature effect and the atomic SOC of carbon.
In Sec.~\ref{sec:SR},
we investigate effects of the effective SOC 
on SR based on the kinetic equation of the spin density operator 
in corrugated SGs,
and calculate 
numerically SRT. 
We give 
a brief summary in Sec.~\ref{sec:Summary}

\section{Effective Spin-orbit coupling}
\label{sec:CISOC}

In this section, we present the effective Dirac Hamiltonian
in flat SGs, and then, the effective SOC Hamiltonian
arising from the local curvature effect combined with the 
atomic SOC in corrugated SGs.
After that, 
we demonstrate how to evaluate 
the effective SOC using the TB Hamiltonian via 
second-order perturbation theory, and
why the SOC can not be inferred from the extension
of that of CNTs.

\subsection{Analytic expression}
\begin{figure}[b!]
\centerline{\includegraphics[width=6cm]{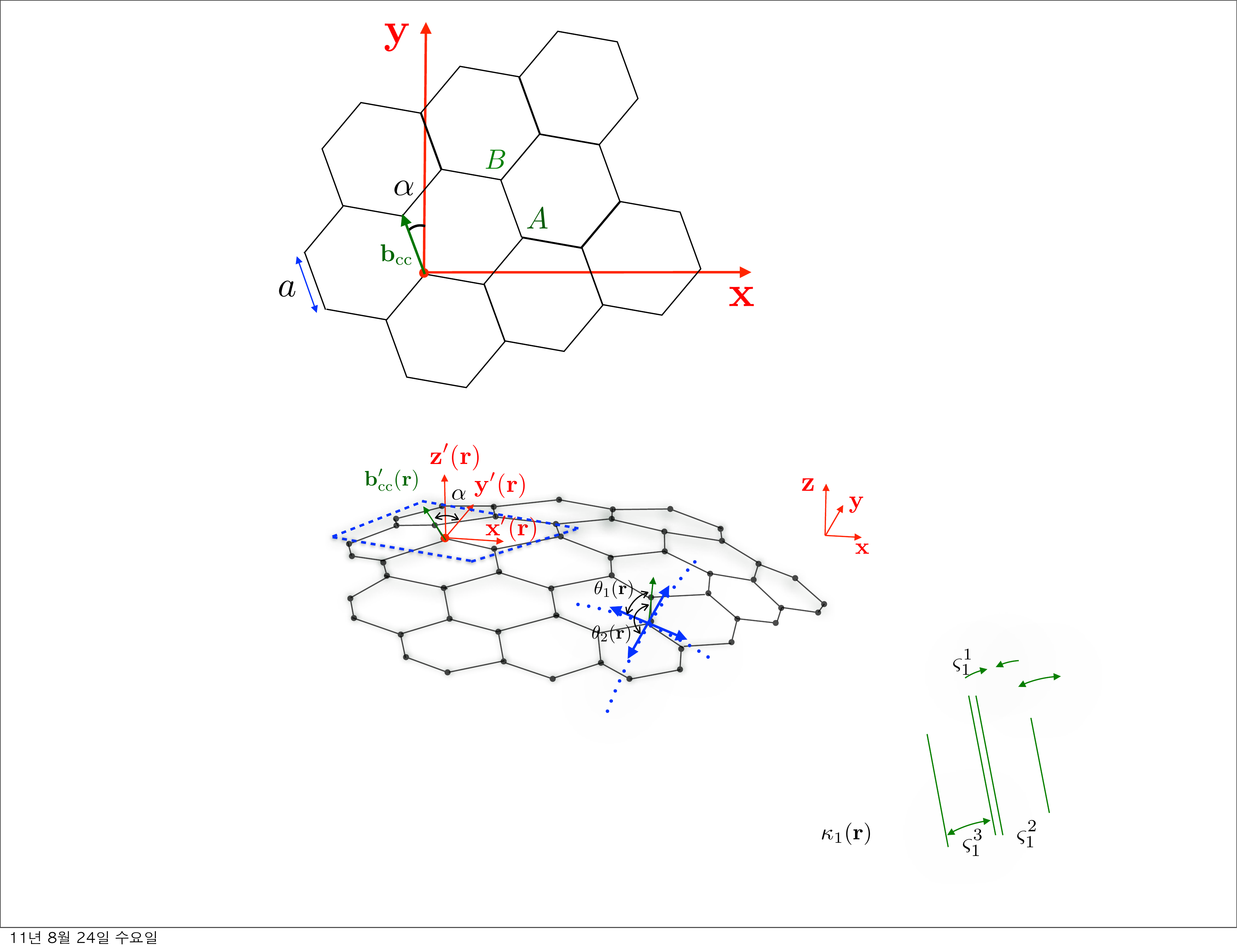}}
\caption{(Color online) Single-layer flat graphene honeycomb lattice with two sublattices $A$ and $B$. 
The angle $\alpha$ is defined counterclockwise from the ${\hat {\bf y}}$ 
to the C-C bond vector ${\bf b}_{\rm cc}$.
$a=|{\bf b}_{\rm cc}|\approx1.42\,{\rm\AA}$ is the carbon-carbon distance.
}
\label{fig:honeycomb}
\end{figure}

Graphene honeycomb lattice with two sublattices $A$ and $B$ (Fig.~\ref{fig:honeycomb}), 
has $\pi$-band consisting of $p_z$ orbitals
near the Fermi level situated at hexagonal corners in the Brillouin zone. 
Close to the Dirac point 
${\bf K}_1$ and ${\bf K}_2\!=\!-{\bf K}_1$,
$\pi$-band states with wave vector ${\bf k}=(k_x, k_y)$ [$|{\bf k}|\ll|{\bf K}_{1(2)}|$] relative to
the ${\bf K}_{1(2)}$, can be described by the effective Dirac Hamiltonian ${\cal H_{\rm Dirac}}$.
When the two sublattices $A$ and $B$ of the honeycomb lattices are used 
as bases, 
the ${\cal H}_{\rm Dirac}$ is written as~\cite{PRB84DiVincenzo}

\begin{equation}
\label{eq:DiracH[1]}
{\cal H}_{\rm Dirac}=
\hbar v_F
\left(
\begin{array}{cc}
0 & e^{i\tau_z\alpha}(\tau_zk_x-ik_y) \\
e^{-i\tau_z\alpha}(\tau_zk_x+ik_y) & 0\\
\end{array}
\right)
\end{equation} 
where $v_F$ is the Fermi velocity,
and $\tau_z=\pm1$ 
denotes the ${\bf K}_{1(2)}$. 
Here, the angle $\alpha$ is defined counterclockwise 
from the ${\hat {\bf y}}$ direction and
the C-C bond vector ${\bf b}_{\rm cc}$-direction
(Fig.~\ref{fig:honeycomb}).
%
%

\begin{figure}[b!]
\centerline{\includegraphics[width=8.5cm]{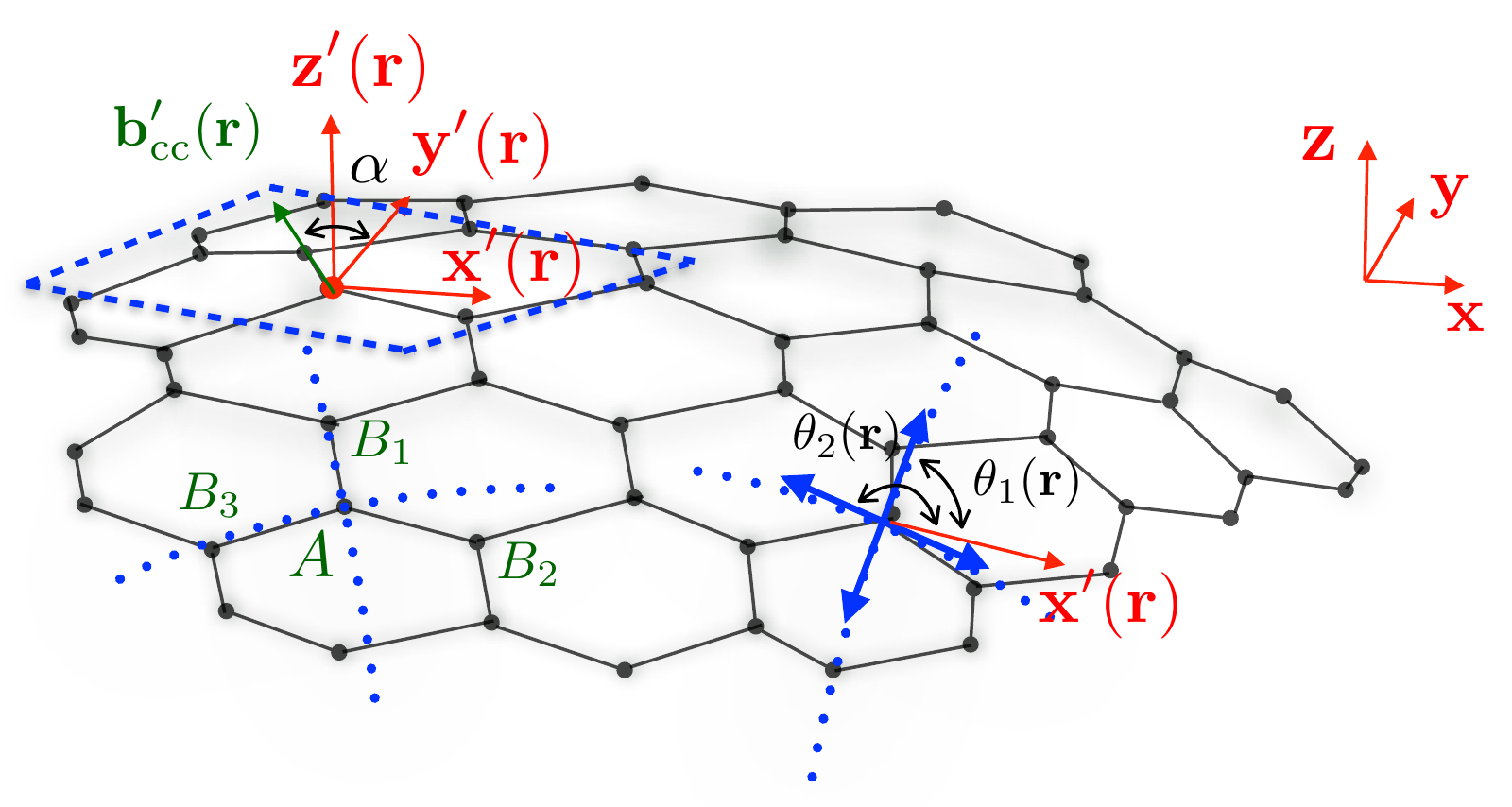}}
\caption{(Color online) Schematic of a partial convex structure in a corrugated  
SG surface described by $z\!=\!h({\bf r})$ where ${\bf r}\!=\!(x, y)$ within the 
2D reference $xy$-plane. 
The
${\hat {\bf z}'}({\bf r})$ 
is the unit vector normal to the local tangential plane
and the ${\bf b}'_{\rm cc}({\bf r})$ (green arrow) is the projected vector from the local C-C bond onto the tangent plane specified by the (blue) dashed quadrangle normal to the ${\hat {\bf z}'}({\bf r})$.
The ${\bf y}'({\bf r})$-direction in the tangent plane is defined  
to make the angle $\alpha$ [Fig.~\ref{fig:honeycomb}] clockwise 
from the ${\bf b}'_{\rm cc}({\bf r})$ in the tangent plane.
Then the ${\bf x}'(\bf r)$-direction is determined automatically
to be mutually orthogonal to ${\bf z}'({\bf r})$-direction and ${\bf y}'({\bf r})$-direction.
Dotted (blue) lines are curvature lines
tangent to principal directions along orthogonal (blue) arrows
where the direction of arrowheads is arbitrary.
The angle $\theta_{1(2)}({\bf r})$ is defined by ${\bf x}'({\bf r})$-direction and the
principal directions, and $\theta_2({\bf r})=\theta_1({\bf r})\pm\pi/2$.
The sublattice $A$ and its nearest-neighbor sites $B_l$ $(l\!=\!1,2,3)$ with two curvature lines
are indicated. 
}
\label{fig:principal curvatures}
\end{figure}

Local curvature effects 
arising from smooth corrugations or ripples 
in SGs
combined with atomic SOC of carbon~\cite{SSC00Serrano}
generate the effective SOC 
for the $\pi$-band states.~\cite{PRB06Huertas-Hernando}
The corrugated SG
can be described as  
a surface $z=h({\bf r})$
with height $h({\bf r})$
as a function of spatial position ${\bf r}=(x, y)$ 
in the two-dimensional (2D) reference plane (Fig.~\ref{fig:principal curvatures}). 
Experiments~\cite{Nature07Meyer,PNAS07Stolyarova,NanoLett07Ishigami,PRL09Geringer,PRB11Knox} suggest
that $h({\bf r})$ shows fluctuations of the order of $1\,{\rm nm}$
over scales $\sim\!10\,{\rm nm}$, 
which can be specified by 
$h_{i}h_j\ll1$ ($i,j=x,y$)
with
$h_{i}$ being the partial derivative along the $i$ direction. 
When the local structure 
of the SG surface 
has two 
principal curvatures $\kappa_1({\bf r})$ and $\kappa_2({\bf r})$,
we find that 
the ${\cal H}_{\rm soc}(\bf r)$ 
is written as 
%
%
%
\begin{align}
\label{eq:CISOC[1]}
{\cal H}_{\rm soc}({\bf r})=
\left(
\begin{array}{cc}
\zeta'\tau_z{\bm \mu}({\bf r})\cdot {\bf s} & \zeta e^{i\tau_z\alpha}\nu({\bf r})(\tau_zs_y+is_x) \\
\zeta e^{-i\tau_z\alpha}\nu({\bf r})(\tau_zs_y-is_x)  & \zeta'\tau_z{\bm \mu}({\bf r})\cdot {\bf s}
\end{array}
\right)
\end{align}
%
%
%
%
where
${\bf s}\!=\!(s_x, s_y, s_z)$
denotes the real spin Pauli matrix.  
Here, 
$\zeta$ and $\zeta'$ are
material parameters, and 
${\bm \mu}({\bf r})$
and $\nu({\bf r})$ are
geometrical parameters containing the local curvature information. 
For $\alpha=0$, the off-diagonal part
of Eq.~(\ref{eq:CISOC[1]})
reduces to the Rashba-type SOC
$\propto(\sigma_y s_x-\tau_z\sigma_x s_y)$
reported in existing theories~\cite{PRB06Huertas-Hernando,PRL09Huertas-Hernando},
where ${\bm \sigma}=(\sigma_x, \sigma_y, \sigma_z)$
denotes the sublattice-pseudospin Pauli matrix. 
On the other hand, the diagonal part
of Eq.~(\ref{eq:CISOC[1]}) is one of main
results of this paper, which was not ignored in previous studies.
Note that this additional SOC term in the diagonal part does not
couple to the pseudospin. 

The material parameters $\zeta$ and $\zeta'$ 
are, respectively, 
given by
$\zeta=a{\lambda_{\rm so}(\epsilon_p-\epsilon_s)(V^{\pi}_{pp}+V^{\sigma}_{pp})
/(12 V^{\sigma2}_{sp})}$
and
$\zeta'=a{\lambda_{\rm so}V^{\pi}_{pp}
/(2(V^{\pi}_{pp}-V^{\sigma}_{pp}))}$,
where $\lambda_{\rm so}$ is the atomic SOC strength of the $p$ orbitals, 
$\epsilon_{s(p)}$ is the atomic energy 
for $s(p)$ orbitals and $a$ is the carbon-carbon distance. 
Here, 
$V^{\sigma}_{sp}$ and $V^{\pi(\sigma)}_{pp}$ represent the coupling
strength in the absence of the curvature for the $\sigma$ coupling between 
nearest-neighbor $s$ and $p$ orbitals and
the $\pi(\sigma)$
coupling between nearest-neighbor $p$ orbitals, respectively. 

The geometrical parameters $\nu({\bf r})$ and ${\bm \mu}({\bf r})$
are given as follows.
$\nu({\bf r})$ is a sum of the two local curvatures, 
$\nu({\bf r})=\kappa_1({\bf r})+\kappa_2({\bf r})$
and ${\bm \mu({\bf r})}$ is a weighted sum of the two local curvatures, 

%
%
\begin{align}
\label{eq:mu}
{\bm \mu}({\bf r})&=\left(\mu_x({\bf r}), \mu_y({\bf r})\right)\nonumber\\
&=\sum_{j=1}^2\kappa_j({\bf r})\left(\sin\!\left[2\theta_j({\bf r})-3\alpha\right], \cos\!\left[2\theta_j({\bf r})-3\alpha\right]\right),
\end{align}
where $\theta_1({\bf r})$ and $\theta_2({\bf r})$ are
the angles between a local ${{\bf x}}'({\bf r})$-direction
and the two principal curvature directions
(see Fig.~\ref{fig:principal curvatures} for their precise definitions).
Since the two principal curvature directions are mutually orthogonal, 
$\theta_2({\bf r})=\theta_1({\bf r})\pm\pi/2$.

Both diagonal and off-diagonal terms 
in Eq.~(\ref{eq:CISOC[1]})
are invariant under the time reversal symmetry. 
The off-diagonal term 
is allowed when the 
mirror symmetry with respect to the $xy$-plane is broken.~\cite{PRL05Kane}
On top of that, owing to the broken $C_3$ symmetry in corrugated SGs,
the diagonal term of the form $\tau_z s_{x(y)}$ is allowed as well.  
Note that the diagonal term has  
the $\theta_{1(2)}({\bf r})$-dependence.
However, 
it can not be obtained from the naive 
extension of the diagonal SOC of CNTs
that depends on
the chiral angle $\theta$
through $\cos(3\theta)$.~\cite{JPSJ09Izumida,PRB09Jeong}
Quantitatively, both terms have the same order of magnitude. 
Using 
$\lambda_{\rm so}\approx 10\,$meV,~\cite{SSC00Serrano} 
$\epsilon_{s}=-7.3\,$eV,
$\epsilon_{p}=0\,$eV,
$V^{\sigma}_{sp}=4.20\,$eV, $V^{\sigma}_{pp}=5.38\,$eV, 
$V^{\pi}_{pp}=-2.24\,$eV in Ref.~\cite{PRL91Tomanek}
we obtain $\zeta'\approx 0.21\,{\rm meVnm}$
and $\zeta\approx 0.15\,{\rm meVnm}$.

The $\theta_{j}({\bf r})$ and the $\alpha$-dependence of ${\cal H}_{\rm soc}({\bf r})$
combined with the SG morphology have interesting implications. 
%
%
First of all, since $\theta_{2}({\bf r})=\theta_1({\bf r})\pm\pi/2$, 
the diagonal term depends on
the difference between two principal curvatures
in contrast to the off-diagonal term depending 
on their sum. 
As a result, at umbilical points where $\kappa_1({\bf r})=\kappa_2({{\bf r}})$,
the diagonal term 
disappears while the off-diagonal term remains, 
which confirms
that every tangent vector   
can be a principal direction 
at the umbilical points.~\cite{Book88Struik} 
In contrast with the umbilical points, 
at saddle points where ${\kappa}_1({\bf r})=-{\kappa}_2({\bf r})$, 
the diagonal term only 
appears while the off-diagonal term disappears.

In addition, 
for periodic ripples having  unidirectional curvature direction 
($\kappa_1({\bf r})\neq 0$ and $\kappa_2({\bf r})=0$), 
the diagonal SOC field direction varies depending on the principal curvature
direction and the honeycomb lattice orientation as well. 
When $\alpha= q/3$ $(q\in \mathbb{Z})$ so that the ${\hat {\bf x}}$ is along the 
zigzag direction [Fig.~\ref{fig:honeycomb}], 
for example, 
the diagonal term is $\propto \kappa_1({\bf r})s_y$
for $\theta_1({\bf r})=0$ and $\pi/2$, 
while it is $\propto\kappa_1({\bf r})s_x$ for $\theta_1({\bf r})=\pi/4$.
On the other hand, when $\alpha =q/3+\pi/6$ so that
the ${\hat {\bf x}}$ is along the armchair direction [Fig.~\ref{fig:honeycomb}], 
the diagonal term is $\propto\kappa_1({\bf r})s_x$
for $\theta_1({\bf r})=0$ and $\pi/2$, while it is $\propto\kappa_1({\bf r})s_y$
for $\theta_1({\bf r})=\pi/4$.
Recent experiments shows that 
the periodic ripple line direction can be controlled~\cite{NatNano09Bao}
and the multiple periodic ripple domains with different ripple lines can occur,~\cite{Science11Choi}
which implies that the variation of the diagonal term depending on the curvature direction
and the honeycomb lattice orientation could be significant 
for the analysis of spin transport in SGs with those structural deformation. 
After all, 
in order to describe precisely 
the curvature-induced SOC in SGs 
both terms should be considered
on an equal footing.

The ${\cal H}_{\rm soc}({\bf r})$ can be 
written as an explicit function of $h({\bf r})$ in the SG surface, $z=h({\bf r})$
where $h_{i}h_{j}\ll1$ ($i,j=x,y$).
The principal curvature 
is the eigenvalue of the shape operator ${\cal S}={\cal F}_1^{-1}{\cal F}_2$
with ${\cal F}_{1(2)}$ being the first (second) 
fundamental form.~\cite{Book88Struik}
Moreover, since its eigenvector ${\hat {\bm v}}_{1(2)}({\bf r})$ 
is along the principal direction, 
$\theta_{1(2)}({\bf r})$
can be determined by the ${\hat {\bm v}}_{1(2)}({\bf r})$ 
and 
${\hat {\bf x}}'({\bf r})$, 
which is assumed as 
${\hat {\bf x}}'({\bf r})\approx(1, 0, h_x)$ in the zeroth order of $h_ih_j$.
${\cal F}_1$ and ${\cal F}_2$
are, respectively, given by~\cite{Book88Struik}

\begin{equation}
\label{eq:fundamentalform}
{\cal F}_1\!=\!
\left(\!
\begin{array}{cc}
1+h_x^2 & h_xh_y\\
h_xh_y & 1+h_y^2 \\
\end{array}
\!\right),\,\,
{\cal F}_2\!=\!
{1\over \sqrt{1+|\nabla h|^2}}
\left(
\begin{array}{cc}
h_{xx} & h_{xy} \\
h_{xy} & h_{yy} \\
\end{array}
\right),
\end{equation}
and then, 
the ${\kappa}_{1(2)}({\bf r})$ 
and ${\hat {\bm v}}_{1(2)}({\bf r})$ are written as 
%
%
${\kappa}_{1(2)}({\bf r})\!\approx\!\big(h_{xx}
+h_{yy}-(-1)^{1(2)}[(h_{xx}-h_{yy})^2
+4h_{xy}^2]^{1\over 2}\big)/2$
%
%
and ${\hat {\bm v}}_{1(2)}({\bf r})\approx
[\alpha_{1(2)}{\hat {\bf x}}+\beta{\hat {\bf y}}+(\alpha_{1(2)}h_x+\beta h_y){\hat {\bf z}}]/{\cal N}$, 
respectively, 
where
$\alpha_{1(2)}\approx h_{xx}-h_{yy}-(-1)^{1(2)}[(h_{xx}-h_{yy})^2+4h_{xy}^2]^{1\over 2}$
and 
$\beta\approx2h_{xy}$ with normalization constant ${\cal N}$.

\subsection{Tight-binding Hamiltonian}

Now we demonstrate 
how to derive Eq.~(\ref{eq:CISOC[1]})
microscopically.
Using second-order perturbation theory~\cite{Book68Schiff}
with   
the 
local curvature 
and 
the atomic SOC 
as weak perturbation, 
we obtain Eq.~(\ref{eq:CISOC[1]}).
Thus we need to know 
the TB Hamiltonian of the local curvature
$H_{\rm c}$ and 
the atomic SOC 
$H_{\rm so}$
taking the honeycomb lattice structure into account exactly.
Most parts of this subsection are devoted to the TB theory of the $\pi$-$\sigma$ hybridization among $s$ and $p$ orbitals.
The unoccupied $d$ orbital effects~\cite{PRB09Gmitra,PRB10Konschuh} are discussed at the end of this subsection.

\begin{figure}[b!]
\centerline{\includegraphics[width=9cm]{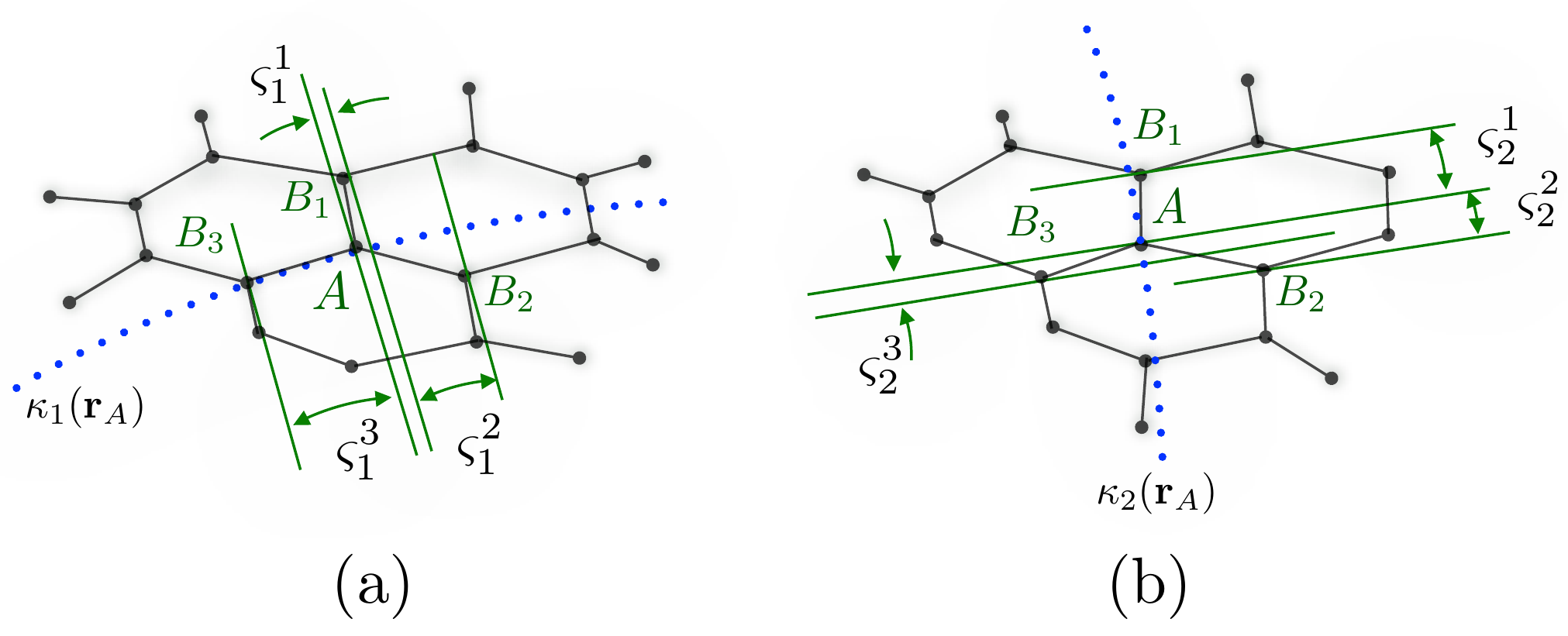}}
\caption{(Color online) Schematic of a partial convex SG surface with sublattices $A$ 
and $B_{l}$ ($l=1,2,3$) indicated in Fig.~\ref{fig:principal curvatures},
which is assumed to have 
unidirectional principal curvature. 
Although the original surface [Fig.~\ref{fig:principal curvatures}] 
has finite 
principal curvatures $\kappa_{1}({\bf r}_A)$ and $\kappa_2({\bf r}_A)$,
each curvature effect is evaluated 
under the assumption that 
the other principal curvature is zero.
$\kappa_1({\bf r}_A)\neq0$ and $\kappa_2({\bf r}_A)=0$ in (a) and
$\kappa_1({\bf r}_A)=0$ and $\kappa_2({\bf r}_A)\neq\!0$ in (b).
Dotted (blue) lines represent curvature lines, and principal direction at ${\bf r}_A$ is tangent to
the curvature lines. 
The arc length $\varsigma^l_{j}$ ($j=1,2$)
between a solid (green) line passing $A$ atom normal to 
the principal direction,
and its parallel solid (green) line passing $B_l$ atom, is related with
$\omega^l_j$ by $\omega^l_{j}\!\approx\!\varsigma^l_{j}\kappa_{j}({\bf r})/2$ [Eq.~(\ref{eq:curvatureTB})],
}
\label{fig:ripple}
\end{figure}

Firstly, 
$H_{\rm c}$
is determined purely by local
curvature effect
dependent on  
$\kappa_{1(2)}({\bf r})$ and $\theta_{1(2)}({\bf r})$
at three bonds between an atomic site
${\bf r}_A$ in the sublattice $A$ and 
nearest-neighbor sites $B_{l}$ ($l=1,2,3$) in the sublattice $B$ [Fig.~\ref{fig:principal curvatures}].
For the smoothly corrugated SGs,
we consider solely one of two principal curvature effects  
assuming that the other principal curvature effect is absent.
There are also mutual curvature effects between them, but 
those effects can be ignored for the smooth corrugated surface
with $h_ih_j\ll1$ $(i,j=x,y)$
since they are higher order in $a/{\cal R}_{1(2)}({\bf r}_A)$
where ${\cal R}_{1(2)}({\bf r}_A)=\kappa^{-1}_{1(2)}({\bf r}_A)$.
Then, it is enough to deal with two unidirectional ripple structures whose principal directions are
orthogonal as shown 
in Fig.~\ref{fig:ripple}.
For the unidirectional ripple structure, 
each principal curvature and the angle between the principal direction and the ${\bf x}'({\bf r})$
at three neighbor C-C bonds around the ${\bf r}_A$ are same 
as $\kappa_{1(2)}({\bf r}_A)$ and $\theta_{1(2)}({\bf r_A})$,
respectively.
Then, up to first order in $a/{\cal R}_{1(2)}({\bf r}_A)$, 
the $H_{\rm c}$
can be approximated as $H_{\rm c}\approx H_{1}+H_{2}$,
each of which describes one principal curvature effect
in the absence of the other one. 
Two principal curvatures yield correlation effects between term in $H_{\rm c}$,
but they are higher order in term of $a/{\cal R}_{1(2)}({\bf r})$ and thus can be ignored. 
The 
$H_{j}$ $(j=1,2)$ is written as~\cite{PRB06Huertas-Hernando,PRB09Jeong}

\begin{align}
\label{eq:curvatureTB}
H_{j}&=\sum_{{\bf r}_A}\sum^3_{l=1}\sum_{\eta=\uparrow,\downarrow}
\Big[
-S^l_{j}\left(c^{s\dagger}_{{\bf r}_A \eta}c^{z'}_{B_l \eta}+c^{z'\dagger}_{{\bf r}_A \eta}c^s_{B_l \eta}\right)\nonumber\\
&+X'^l_{j}\Big(c^{x'\dagger}_{{\bf r}_A \eta}c^{z'}_{B_l \eta}-c^{z'\dagger}_{{\bf r}_A \eta}c^{x'}_{B_l \eta}\Big)\nonumber\\
&+Y'^l_{j}\Big(c^{y'\dagger}_{{\bf r}_A \eta}c^{z'}_{B_l \eta}-c^{z'\dagger}_{{\bf r}_A \eta}c^{y'}_{B_l \eta}\Big)\Big]
+{\rm H.c.},
\end{align}
where
$c_{r\uparrow(\downarrow)}^{x'}$, $c_{r\uparrow(\downarrow)}^{y'}$, and $c_{r\uparrow(\downarrow)}^{z'}$ represent the annihilation operators for
$p_{x'}$, $p_{y'}$, $p_{z'}$ orbital states with eigenspinor $\chi_{\uparrow(\downarrow)}$
along $\hat{\bf z}'$ at a carbon atom $r$.
Here,  
$S^l_{j}$, $X'^l_{j}$, and $Y'^l_{j}$ represent 
the curvature-induced coupling strength of $p_{x'}$, $p_{y'}$, and $s$ orbitals
with a nearest-neighbor $p_{z'}$ orbitals.
Their expressions are written as $S^{l}_{j}\!=\!\omega_{j}^l\tilde{S}^l_{j}$,
$X'^l_{j}\!=\!\omega^l_{j}\tilde{X}^l_{j}\!\cos[\theta_{j}({\bf r}_A)]\!-\!\omega^l_{j}\tilde{Y}^l_{j}\sin[\theta_{j}({\bf r}_A)]$, 
and
$Y'^l_{j}\!=\!\omega^l_{j}\tilde{X}^l_{j}\sin[\theta_{j}({\bf r}_A)]\!+\!\omega^l_{j}\tilde{Y}^l_{j}\cos[\theta_{j}({\bf r}_A)]$,
with $\omega^1_{j}\!\approx\! {a/(2{\cal R}_{j}({\bf r}_A))}\sin[\theta_{j}({\bf r}_A)-\alpha]$, 
$\omega^2_{j}\!\approx\! a/(2{\cal R}_{j}({\bf r}_A))\sin[\pi/3-\theta_{j}({\bf r}_A)+\alpha]$,
and $\omega^3_j\!\approx\!a/(2{\cal R}_j({\bf r}_A))\sin[\pi/3+\theta_{j}({\bf r}_A)-\alpha]$.
Here, $\tilde{S}^l_{j}$, $\tilde{X}^l_j$, and $\tilde{Y}^l_j$
are, respectively, written as

\begin{align}
\label{eq:SXY}
\tilde{S}^1_j&=V^{\sigma}_{sp}\sin\left[\beta_j({\bf r}_A)\right],\nonumber\\
\tilde{S}^2_j&=V^{\sigma}_{sp}\cos\left[{\pi\over 6}+\beta_j({\bf r}_A)\right],\nonumber\\
\tilde{S}^3_j&=V^{\sigma}_{sp}\cos\left[{\pi\over 6}-\beta_j({\bf r}_A)\right],\nonumber\\
\tilde{X}^1_j&=-V^{\pi}_{pp}-V^{\pi}_{pp}\cos^2\left[\beta_j({\bf r}_A)\right]-V^{\sigma}_{pp}\sin^2\left[\beta_j({\bf r}_A)\right],\nonumber\\
\tilde{X}^2_j&=-V^{\pi}_{pp}-V^{\pi}_{pp}\cos^2\left[{\pi\over 3}\!-\beta_j({\bf r}_A)\right]\!-V^{\sigma}_{pp}\sin^2\!\left[{\pi\over 3}\!-\beta_j({\bf r}_A)\right],\nonumber\\
\tilde{X}^3_j&=V^{\pi}_{pp}+V^{\pi}_{pp}\cos^2\left[{\pi\over 6}-\beta_j({\bf r}_A)\right]+V^{\sigma}_{pp}\sin^2\left[{\pi\over 6}-\beta_j({\bf r}_A)\right],\nonumber\\
\tilde{Y}^1_j&={1\over 2}\left(V^{\pi}_{pp}-V^{\sigma}_{pp}\right)\sin\left[2\beta_j({\bf r}_A)\right],\nonumber\\
\tilde{Y}^2_j&={1\over 2}\left(V^{\pi}_{pp}-V^{\sigma}_{pp}\right)\sin\left[2\beta_j({\bf r}_A)-{2\pi\over 3}\right],\nonumber\\
\tilde{Y}^3_j&={1\over 2}\left(V^{\pi}_{pp}-V^{\sigma}_{pp}\right)\sin\left[2\beta_j({\bf r}_A)-{\pi\over 3}\right],
\end{align}
%
%
%
%
where $\beta_{j}({\bf r}_A)\!=\!\theta_{j}({\bf r}_A)-\alpha$.
Note that the local curvature 
generates
interband transition between $\pi$-band and 
$\sigma$-band consisting of $s$, $p'_x$, and 
$p'_y$ orbitals.~\cite{PRB06Huertas-Hernando,EPL08Kim,PRB09Jeong}

Secondly, 
$H_{\rm so}$
is given by  
$H_{\rm so}\!=\!\lambda_{\rm so}\sum_{r}{\bf L}_{r}\cdot{\bf S}_{r}$,~\cite{PRB06Huertas-Hernando,PRB06Min}
where ${\bf L}_r({\bf S}_r)$ is
atomic-orbital (spin) angular momentum of an electron
at $r$,
and can be expressed by~\cite{PRB06Huertas-Hernando,PRB09Jeong} 

%
%
\begin{align}
\label{eq:atomicTB}
H_{\rm so}=&{\lambda_{\rm so}\over 2}\sum_{r={\bf r}_{A/B}}
\Big(c_{r\downarrow}^{z'\dagger}c_{r\uparrow}^{x'}-c_{r\uparrow}^{z'\dagger}c_{r\downarrow}^{x'}+ic_{r\uparrow}^{z'\dagger}c_{r\downarrow}^{y'}+ic_{r\downarrow}^{z'\dagger}c_{r\uparrow}^{y'}\nonumber\\
&+ic_{r\uparrow}^{y'\dagger}c_{r\uparrow}^{x'}-ic_{r\downarrow}^{y'\dagger}c_{r\downarrow}^{x'}\Big)+{\rm H.c.}.
\end{align}
Note that on-site coupling between $p_{z'}$ and $p_{x'(y')}$ orbitals 
generates $\pi$-$\sigma$ interband 
transition that, combined with the curvature-induced interband transition,
contributes to the ${\cal H}_{\rm soc}({\bf r})$.

In CNTs, however, the coupling between 
$p_{z'}$ orbital and $p$ orbital along the CNT axis has no contribution
to the effective SOC. 
For instance,
if the Eq.~(\ref{eq:atomicTB}) is expressed
with eigenspinor along the CNT axis parallel 
to ${\hat {\bf y}}'({\hat {\bf x}}')$-direction instead of $\chi_{\uparrow(\downarrow)}$,
the coupling between $p_{z'}$ and $p_{y'(x')}$ orbitals
acquires a phase term $e^{\pm i\phi}$ with the azimuthal angle $\phi$  
along the circumference~\cite{PRB06Huertas-Hernando,PRB09Jeong},
and, as a result, its contribution averages out to zero after integration 
over the circumference.~\cite{PRB06Huertas-Hernando}
Hence, 
the trivial extension from the curvature-induced SOC of the CNT to that of the SG is 
inevitably incomplete
because of their topological difference in the geometrical structure.

Second-order process resulting from two consecutive 
$\pi$-$\sigma$ interband transitions
gives rise to
the effective SOC Hamiltonian
for the unperturbed $\pi$-band states of Eq.~(\ref{eq:DiracH[1]}).~\cite{Second-order perturbation}
For the $\sigma$-band calculation, we use the Slater-Koster parametrization~\cite{PR54Slater}
of $s$, $p_x$, and $p_y$ orbitals.
During the second-order process, pseudospin-conserving 
and pseudospin-flipping processes occur;
the former and the latter generate, respectively, 
the diagonal and the off-diagonal SOC in ${\cal H}_{\rm soc}({\bf r})$ [Eq.~(\ref{eq:CISOC[1]})].~\cite{PRB09Jeong}
%
The resulting effective SOC, then,
is expressed in local coordinate axes of
${\hat {\bf x}}'({\bf r})$, ${\hat {\bf y}}'({\bf r})$, ${\hat {\bf z}}'({\bf r})$. 
However,
for the smooth corrugation where
$h_{i}h_{j}\ll1$ $(i,j=x,y)$,
${\hat {\bf x}}'({\bf r})$, ${\hat {\bf y}}'({\bf r})$, ${\hat {\bf z}}'({\bf r})$
in the resulting effective SOC may be replaced 
by ${\hat {\bf x}}$, ${\hat {\bf y}}$, ${\hat {\bf z}}$ to obtain 
${\cal H}_{\rm soc}({\bf r})$ in Eq.~(\ref{eq:CISOC[1]})
since the difference between the two sets of 
axes [for instance, 
${\hat {\bf z}'}({\bf r})\cdot{\hat {\bf z}}
=\left(1+|\nabla h({\bf r})|^2\right)^{-{1\over 2}}\approx 1$]
generates higher order terms in $a/{\cal R}_{1(2)}({\bf r})$.

Lastly, we remark $d$ orbitals effects on the curvature-induced SOC,
which were not considered in our calculation.
For the spin-orbit induced gap, 
the $d$ orbital effects are recently reported to be
important since the hybridization between 
$d$ orbitals and $p$ orbitals 
generate the gap, 
which is linear in the atomic SOC strength $\lambda_{d}$ 
of $d$ orbitals.~\cite{PRB09Gmitra,PRB10Konschuh}
This linear contribution can be dominant over the $p$ orbital contribution,
which is quadratic in $\lambda_{\rm so}$. 
In contrast, the contribution of the $d$ orbitals to the external field-induced Rashba SOC
is smaller than that of the $p$ orbitals.~\cite{PRB10Konschuh}
The coupling between $p$ and $d$ orbitals can also lead to the curvature-induced SOC
in the first order of $a/{\cal R}_{1(2)}({\bf r})$.
Due to the symmetrical reason, however,
its form
is expected to be same as the form of ${\cal H}_{\rm soc}({\bf r})$ [Eq.~(\ref{eq:CISOC[1]})]
except parameters in ${\zeta}$ and ${\zeta'}$.
The curvature-induced SOC owing to the $d$ orbitals can be generated as follows.
The $\pi$-band
consisting of the $p_z$ orbital and the nearest-neighbor $d_{xz}$ and $d_{yz}$ 
orbitals
can couple with  the $\sigma$ bands through on-site coupling between 
$d$ orbitals in $\pi$-band and $\sigma$-band by the atomic SOC.
Then, electrons in $\sigma$-band can
delocalize to the nearest-neighbor $p$ and $d$ orbitals within $\sigma$ bands. 
Finally, the $p$ and $d$ orbitals 
in $\sigma$-bands
can couple with the nearest-neighbor $p_z$ orbitals due to the curvature, 
yielding the curvature-induced SOC
that is proportional to 
$a/{\cal R}_{1(2)}({\bf r})$, 
$\lambda_{d}$,
and $(\epsilon_{d}-\epsilon_{p})^{-2}$.
Based on the estimation of the Rashba SOC strength
and comparison with the $p$ orbital contribution given in Ref.~\cite{PRB10Konschuh},
we can compare roughly the $d$ orbital contribution of the curvature-induced SOC 
with the ${\cal H}_{\rm soc}({\bf r})$ [Eq.~\ref{eq:CISOC[1]}].
The $\lambda_{d}$ is smaller than $\lambda_{\rm so}$, 
and further, $(\epsilon_{d}-\epsilon_{p})^{-2}$ multiplied by hopping parameters associated with
$d$ and $p$ orbitals is a small quantity
compared with 
parameters in same dimension in $\zeta$ and $\zeta'$.
Hence, we believe that the contribution of $d$ orbitals 
to the curvature-induced SOC is smaller than that of $p$ orbitals.
In order to get more reliable quantitative estimates, 
systematic study considering coupling between $p$, $s$ orbitals and  
$d$ orbitals is required.~\cite{PRB10Konschuh}


\section{Spin relaxation}
\label{sec:SR}
In this section, 
we investigate effects of the geometric curvature
on SR in chemically clean SGs with
nanoscale corrugations. 
Motivated by recent experiments~\cite{NanoLett07Ishigami}
about the corrugation roughness of SGs,
we study SR dependence on the fractal dimension
of the corrugation roughness
within the assumption that both SOC and momentum scattering~\cite{PTTSA08Katsnelson}
arise mainly from the corrugation.  
Since the substrate-induced SOC effect on SR was addressed
in a previous theory,~\cite{PRB09Ertler} we focus on the curvature-induced SOC here.

\subsection{Kinetic equation of the real spin density opeator}

In order to calculate SRT,
we use
the kinetic equation
of the real spin density operator $\rho$.~\cite{PRB11Dugaev,JPhys02Averkiev}
Besides the ${\cal H}_{\rm soc}({\bf r})$, 
we include
the strain-induced vector potential ${\cal H}_{\rm v}({\bf r})$
that occurs in corrugated SGs.~\cite{JPSJ06Ando,PRL06Morozov,PRB08Guinea}
Although ${\cal H}_{\rm v}({\bf r})$ is independent of the spin, 
it is still relevant for the SRT
since it affects the momentum scattering rate.  
Then the local potential Hamiltonian ${\cal H}_{\rm p}({\bf r})$ is written as 
${\cal H}_{\rm p}({\bf r})\!\equiv\!{\cal H}_{\rm soc}({\bf r})+{\cal H}_{\rm v}(\bf r)$
which is considered 
as weak perturbation in the kinetic equation. 
We assume
$\langle {\cal H}_{\rm p}({\bf r})\rangle\approx 0$
where
$\langle\cdots\rangle$ denotes ensemble average.
Since ${\cal H}_{\rm p}({\bf r})$ is expected to change slowly
over scales larger than the lattice spacing so that 
the hybridization between $\tau_z\!=\!1$ and $\tau_z\!=\!-1$ states
can be ignored, 
we calculate SRT for the electron state of the ${\cal H}_{\rm Dirac}$ [Eq.~(\ref{eq:DiracH[1]})] 
when $\tau_z\!=\!1$ and $\alpha\!=\!0$, 
where  eigenenergy $\varepsilon_{\bf k}$
and eigenstate $|\psi_{\bf k}\rangle$ are, respectively, given by 
$\varepsilon_{\bf k}\!=\!\hbar v_F{k}$ with $k\!\equiv\!|{\bf k}|$, and 
$|\psi_{\bf k}\rangle\!=\!1/\sqrt{2}(e^{-i\varphi_{\bf k}}~1)^{\rm T}|{\bf k}\rangle$ 
with polar angle $\varphi_{\bf k}$ of the wavevector ${\bf k}$.~\cite{electron_hole_symmetry}
Then a $2\!\times\!2$ real spin density operator $\rho_{\bf k}$ diagonal in momentum
has a representation in eigenspinors chosen to be
along 
${\hat {\bf z}}$ direction.
The kinetic equation of $\rho_{\bf k}$
is written as~\cite{PRB11Dugaev,JPhys02Averkiev}
\begin{align}
\label{eq:kineticEq[1]}
&{\partial {\rho}_{\bf k}\over \partial t}
+{i\over \hbar}\left[V_{\bf kk}, {\rho}_{\bf k}\right]={\cal C}_{{\bf k}},
\end{align}
where
$V_{\bf kk'}\!=\!\langle \psi_{\bf k}|{\cal H}_{\rm p}({\bf r})|\psi_{\bf k'}\rangle$,
and ${\cal C}_{\bf k}$ is the collision integral describing 
momentum scattering, which is given by
${\cal C}_{\bf k}\!=\!{\pi/\hbar} \sum_{{\bf k}'}
\left[2V_{{\bf k}{\bf k}'}{\rho}_{{\bf k}'}V_{{\bf k}'{\bf k}}
-\left\{{\rho}_{{\bf k}}, V_{{\bf k}{\bf k}'}V_{{\bf k}'{\bf k}}\right\}\right]
\delta(\varepsilon_{{\bf k}'}-\varepsilon_{\bf k})$.
The commutator in Eq.~(\ref{eq:kineticEq[1]})
can be ignored because 
${\cal H}_{\rm p}({\bf r})$ has no regular contribution.
%
%
%
Since the energy scale of ${\cal H}_{\rm v}({\bf r})$
is typically larger than that of ${\cal H}_{\rm soc}({\bf r})$,
the SRT would be longer than a time scale in which the momentum
distribution during momentum scattering becomes isotropic.~\cite{JPhys02Averkiev}
Thus, 
the ${\rho}_{\bf k}$ can be represented as
${\rho}_{\bf k}\!=\!{\overline {\rho}}+{\rho}'_{\bf k}$
with the anisotropic part $\rho'_{\bf k}$ due to the ${\cal H}_{\rm soc}({\bf r})$,
where
the bar means averaging over $\varphi_{\bf k}$. 
Inserting the $\rho_{\bf k}\!=\!\overline{\rho}+\rho_{\bf k}'$ into Eq.~(\ref{eq:kineticEq[1]}), 
we obtain the equation of $\overline{\rho}$ as, 
\begin{align}
\label{eq:kineticEq[2]}
{\partial \overline{\rho}\over \partial t}\!=\!{\pi d\over \hbar}\!
\!\oint\!d\varphi_{\bf k}\!\oint\!d\varphi_{\bf k'}
\!\!\left(\!V'_{\bf kk'}\overline{\rho}V'_{\bf k'k}-
{1\over 2}\left\{\overline{\rho},\left|V'_{\bf kk'}\right|^2\right\}\!\right),
\end{align}
where $d\!=\!k/(\pi \hbar v_F)$ is the density of states. 
Note that collision term consisting of the product of 
${\cal H}_{\rm soc}({\bf r})$ and ${\cal H}_{\rm v}(\bf r)$
disappears  since there is no correlation between them. 
More specifically, since the vector potential is written as quadratic 
in $h_{i,j}$,~\cite{JPSJ06Ando,PRL06Morozov}
the collision terms proportional to 
$\langle h_{i}h_{j}{\bm \mu}({\bf r})\rangle$ or
$\langle h_{i}h_{j}{\nu}({\bf r})\rangle$
vanish by 
the Wick theorem.~\cite{PTTSA08Katsnelson}
Thus, the random gauge field does not contribute to the SR in chemically clean SGs.~\cite{no correlation}

Since the real spin density 
${\bf P}\!=\!(P_x, P_y, P_z)$
can be evaluated by 
${\bf P}\!=\!{\rm Tr}\left(\overline{\rho}{\bf s}\right)$,
we can obtain the kinetic equation of ${\bf P}$ by tracing  Eq.~(\ref{eq:kineticEq[2]})
multiplied by ${\bf s}$
over the real spin space,
which is written as
$\partial P_{i}/\partial t\approx-P_{i}/\tau_i^s$ 
$(i=x,y,z)$
with $\tau_{i}^s$ being the SRT of spin along the $i$-direction. 
Here, 
$\tau^s_{x(y)}$ and $\tau^s_z$ 
are evaluated as
\begin{widetext}
\begin{equation}
\label{eq:SRT[1]}
{1\over\tau^s_{x(y)}}
={4{\pi}^{3}d\over \hbar}\!\int\!\int d{\bf r}d{\bf r'}
\left[\zeta'^2\langle\mu_{y(x)}({\bf r})\mu_{y(x)}({\bf r'})\rangle{J} 
+\zeta^2\langle\nu({\bf r})\nu({\bf r'})\rangle J'\right],
\end{equation}
\end{widetext}
%
%
and ${1/\tau^s_z}=1/\tau^s_x+1/\tau^s_y$, 
where
$J$ and $J'$ are, respectively, 
defined as 
$J\equiv J^2_0({k}R)+J^2_1({k}R)$,
and 
$J'\equiv J^2_0({k}R)$
with $J_{0(1)}(x)$ being the Bessel function of zeroth (first) order and
$R=|{\bf r}-{\bf r'}|$. 
If we assume $\langle\mu_x({\bf r})\mu_x({\bf r'})\rangle\approx
\langle\mu_y({\bf r})\mu_y({\bf r'})\rangle$
and 
$\langle\mu_x({\bf r})\mu_y({\bf r'})\rangle\approx0$
for the random corrugation, 
we obtain $\tau_{x}^s\approx\tau_{y}^s$ and $\tau^s_{x(y)}\approx2\tau^s_z$,
which means that spins out of plane relax twice 
as fast as spins in plane.~\cite{PRL09Huertas-Hernando} 
Note that since the carrier density $n$ is given by $n=k^2/\pi$, 
the SR vanishes at the Dirac points where $n=0$.~\cite{PRB11Dugaev}

\subsection{Numerical results}

\begin{figure}[b!]
\centerline{\includegraphics[width=8cm]{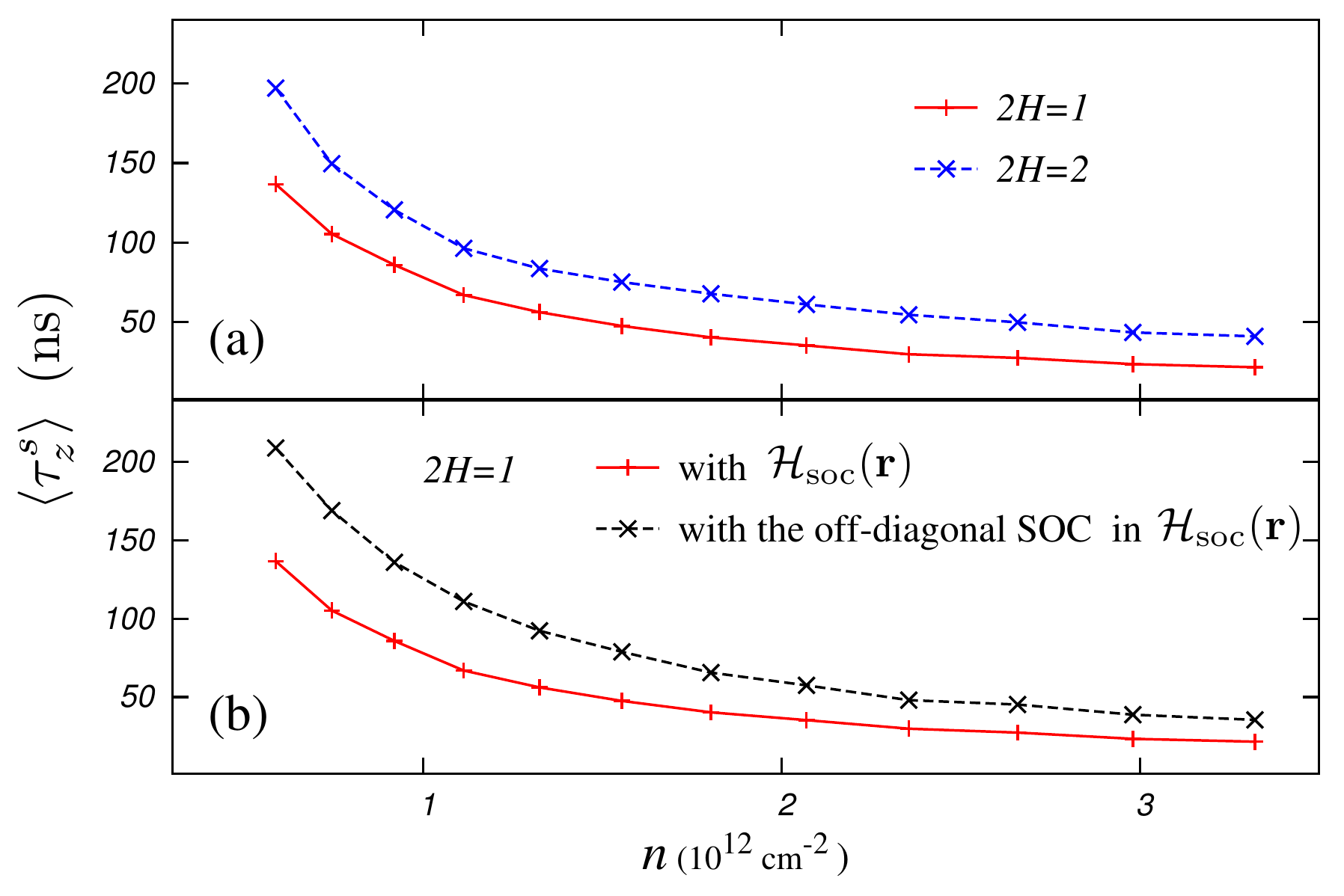}}
\caption{(Color online) 
Spin relaxation time $\langle\tau^s_z\rangle$ 
as a function of the carrier density $n\!=\!{k}^2/\pi$
and the fractal dimension $H$.
The $\langle\tau^{s}_z\rangle$ 
are ensemble-averaged quantities over ten surfaces for $\alpha=0$. 
The corrugated square lattice 
with the height $h({\bf r})$,  the 
lattice constant of  $0.2\,$nm, 
and the side length of $200\,$nm,
are constructed numerically based on the random midpoint 
displacement method~\cite{Communications82Fournier}
with   
height-height correlation function 
$g(R)\!=\!
2\gamma^2\big(1-e^{-({R/\xi})^{2H}}\big)$. 
Here, we set $\gamma\!=\!0.3\,$nm and $\xi\!=\!20\,{\rm nm}$
from recent experiments,~\cite{NanoLett07Ishigami,PRL09Geringer}
and interpolate discrete $h({\bf r})$
in order to
mimic the continuous SG surface.
(a) $\langle\tau^s_z\rangle$ as a function of $n$ for $2H=1$ (red solid line) and $2H=2$ (blue dashed line).
(b) $\langle\tau^s_z\rangle$ as a function of $n$ for $2H=1$
considering ${\cal H}_{\rm soc}({\bf r})$ (red solid line)
and only the off-diagonal SOC in ${\cal H}_{\rm soc}({\bf r})$ (black dashed line), respectively.
}
\label{fig:SRT}
\end{figure}

Figure~\ref{fig:SRT} shows SRTs, $\langle\tau^s_z\rangle$ that are calculated 
numerically in corrugated SG surfaces for $\alpha\!=\!0$
and 
are ensemble-averaged for ten surfaces.
Those surfaces are approximated as square lattices, 
which can be justified since
the ${\cal H}_{\rm soc}({\bf r})$ contains the curvature effects 
taking into account the honeycomb lattice. 
The surface with fluctuating $h({\bf r})$
is constructed based on the 
height-height correlation function
$g(R)\!=\!\langle (h({\bf r})-h({\bf r'}))^2\rangle\!=\!2\gamma^2\big(1-e^{-({R/\xi})^{2H}}\big)$,
where
$H$ characterizes 
the fractal dimension,~\cite{NanoLett07Ishigami,PTTSA08Katsnelson,PRB93Palasantzas}
$\xi$ represents the correlation length,
and $\gamma\!=\!\sqrt{{\langle h^2({\bf r})\rangle}}$.
As the fractal dimension $H$ increases,
the corrugated surface becomes smooth.

A recent experiment~\cite{Nature07Meyer} in suspended SGs shows that
the height of corrugations grows approximately linearly with increasing length, 
which corresponds to a corrugated surface for $2H\!=\!2$. 
For exfoliated SGs on the substrate, on the other hand,
the fractal dimension with $2H\!\approx\!1$ was observed.~\cite{NanoLett07Ishigami}
There is a controversy 
over the origin of observed corrugation roughness.~\cite{PRL10Cullen}
Here, 
we investigate SR in corrugated SGs for $2H\!=\!1$ and $2H\!=\!2$
representatively.
As shown in Fig.~\ref{fig:SRT} (a),
the $\langle\tau^s_z \rangle$ becomes shorter with decreasing $2H$, 
which can be understood in terms of 
a growing irregularity of ${\cal H}_{\rm soc}({\bf r})$ [Eq.~\ref{eq:CISOC[1]}] induced by rougher surface corrugation. 
Also, we check that the $\langle\tau^s_{x(y)}\rangle$ is approximately twice as large as the $\langle\tau^s_{z}\rangle$ (not shown here). 

Next, 
in order to investigate the effect of the diagonal SOC in ${\cal H}_{\rm soc}({\bf r})$, 
we compare two SRTs,
one calculated considering the entire ${\cal H}_{\rm soc}({\bf r})$
and the other considering only the off-diagonal SOC term in ${\cal H}_{\rm soc}({\bf r})$, 
respectively, in SGs with $2H\!=\!1$.
As shown in Fig.~\ref{fig:SRT} (b),
the latter SRT is noticeably larger than the former true SRT. 
Hence, it is necessary 
to consider both SOC terms
for the precise analysis of SR depending on the corrugation roughness.

The calculated SRT as shown in Fig.~\ref{fig:SRT}
is at least of the order $\sim\!10\,{\rm ns}$,
and 
becomes shorter monotonically 
as the carrier density $n$ increases,
which is primarily because
the density of states $d$ increases linearly with $n^{1/2}$ [Eq.~(\ref{eq:SRT[1]})].
We check that the integrand of Eq.~(\ref{eq:SRT[1]}) has little dependence relatively upon $k$.
This implies that the effective spatial range of the random SOC affecting the SR substantially 
does not depend significantly on the variation of $k$ over the scale $\sim\!10\,{\rm nm}^{-1}$.
Our quantitative and qualitative results 
are in agreement with a recent theory.~\cite{PRB11Dugaev}
As mentioned in the Ref.~\cite{PRB11Dugaev}, 
however, those results cannot explain recent spin transport experiments.~\cite{Nature07Tombros}

One possibility for that discrepancy
is that the effect of charged impurity in the substrate 
causing momentum scattering~\cite{JPSJ07Ando}
is not considered in our calculation. 
The momentum scattering  
combined with the curvature-induced Rashba-type SOC
was already investigated theoretically,~\cite{PRL09Huertas-Hernando}
but it cannot explain the measured SRT, either.  
However, the inclusion of the additional diagonal SOC in ${\cal H}_{\rm soc}({\bf r})$
in addition to the off-diagonal SOC
into the EY mechanism
could give a possibility to calculate the SRT 
more precisely. 
In the EY mechanism,~\cite{RMP04Zutic}
spin flip  scattering occurs 
at the very moment when the momentum scattering takes place
since the electron wave functions normally
have an admixture of the opposite-spin states due to the SOC.




\section{Conclusion}
\label{sec:Summary}

We have demonstrated that
the combined action of the curvature and the atomic SOC of carbon 
gives rise to two types of the effective SOC in SGs with corrugation or periodic ripples. 
One of them couples with the sublattice pseudospin,
which corresponds to the 
Rashba-type SOC in a corrugated SG
reported in previous theories,~\cite{PRB06Huertas-Hernando,PRL09Huertas-Hernando, PRB10Zhou,PRB11Dugaev}. 
The other SOC, on the other hand, 
does not couple with the pseudospin,
and was not recognized in previous literature. 
The additional curvature-induced 
SOC depends on the principal curvature direction,
which is similar to the curvature-induced SOC of CNTs
whose diagonal term in pseudospin representation 
depends on the chiral angle of CNTs. 
However, the curvature-induced SOC in SGs can not be obtained from the trivial 
extension
of the curvature-induced SOC in CNTs due to their distinct topological structure between the SG surface
and the CNT cylinder.

We have also investigated SR in chemically clean SGs with nanoscale 
corrugation, 
and found that the diagonal SOC makes the same order of contribution 
to SRT as the off-diagonal SOC. 
The SRT becomes longer as
the fractal dimension of the corrugation roughness increases
since the irregularity of the SOC decreases in smoother SGs. 
The calculated SRT in chemically clean SGs, however,  
can not explain recent experimental results of SR in current exfoliated SG samples. 

A natural direction for future research would be to calculate SRT
in the presence of charged impurities that cause momentum scattering,~\cite{JPSJ07Ando}
considering both the diagonal and the off-diagonal SOC. 
%
In addition to the SR in SGs,   
we expect that
the curvature-induced SOC $[{\cal H}_{\rm soc}({\bf r})]$
may be relevant for the analysis of other spin-related phenomena in SGs.

\begin{acknowledgements}
We thank D. Gang, T. Takimoto, S. Kettemann, K. Kim, B. H. Kim, J. Han, 
I. Kim, S. Kim, K. Hashimoto, F. Ziltener, Y.-W. Son, S.-M. Choi, and S. Lee for useful discussions.
This work was financially supported by the NRF (2010-0014109, 2011-0030790) and BK21.
\end{acknowledgements}

%



\begin{thebibliography}{99}


\bibitem{RMP09Neto}
A. H. Castro Neto, F. Guinea, N. M. R. Peres, K. S. Novoselov, and
A. K. Geim, Rev. Mod. Phys. {\bf 81}, 109 (2009);
A. K. Geim, Science {\bf 324}, 1530 (2009).











\bibitem{Nature06Son}
Y.-W. Son, M. L. Cohen, and S. G. Louie, Nature (London) {\bf 444}, 347 (2006).
W. Y. Kim and K. S. Kim, Nat. Nanotechnol. {\bf 3}, 408 (2008). 

\bibitem{NaturePhys07Trauzettel}
B. Trauzettel, D. V. Bulaev, D. Loss, and G. Burkard, Nat. Phys. {\bf 3}, 192 (2007);
P. Recher and B. Trauzettel, Nanotechnology {\bf 21}, 302001 (2010).




\bibitem{Nature07Tombros}
N. Tombros, C. Jozsa, M. Popinciuc, H. T. Jonkman, and B. J. van Wees, Nature (London)
{\bf 448}, 571 (2007);
N. Tombros, S. Tanabe, A. Veligura, C. Jozsa, M. Popinciuc, H. T.
Jonkman, and B. J. van Wees, Phys. Rev. Lett. {\bf 101}, 046601
(2008);
C. J\'{o}zsa, T. Maassen, M. Popinciuc, P. J. Zomer, A. Veligura, H. T. Jonkman, and 
B. J. van Wees, Phys. Rev. B {\bf 80}, 241403(R) (2009);
W. Han, and R. K. Kawakami,  Phys. Rev. Lett. {\bf 107}, 047207 (2011);
S. Jo, D.-K. Ki, D. Jeong, H.-J. Lee, and S. Kettemann, 
Phys. Rev. B {\bf 84}, 075453 (2011);
W. Han, K. M. McCreary, K. Pi, 
W. H. Wang, Y. Li, H. Wen, J. R. Chen, and R. K. Kawakami, J. Magn. Magn. Mater. {\bf 324}, 369 (2012).



\bibitem{Nature02Jedema}
F. J. Jedema, H. B. Heersche, A. T. Filip, J. J. A. Baselmans, and B. J. van Wees, 
Nature (London) {\bf 416}, 713 (2002).

\bibitem{PRB03Jedema}
F. J. Jedema, M. S. Nijboer, A. T. Filip, and B. J. van Wees, 
Phys. Rev. B {\bf 67}, 085319 (2003).


\bibitem{RMP04Zutic}
I. \v{Z}uti\'c, J. Fabian, S. Das Sarma, Rev. Mod. Phys. {\bf 76}, 323 (2004).

\bibitem{PRL09Neto}
A. H. Castro Neto and F. Guinea, 
Phys. Rev. Lett. {\bf 103}, 026804 (2009). 



\bibitem{PRB09Ertler}
C. Ertler, S. Konschuh, M. Gmitra, and J. Fabian, 
Phys. Rev. B {\bf 80}, 041405 (R) (2009).

\bibitem{PRB06Huertas-Hernando}
D. Huertas-Hernando, F. Guinea, and A. Brataas, 
Phys. Rev. B {\bf 74}, 155426 (2006).



\bibitem{arXiv11Ochoa}
P. Zhang and M. W. Wu, Phys. Rev. B {\bf 84}, 045304 (2011);
H. Ochoa, A. H. Castro Neto, and F. Guinea, 
arXiv:1107.3386v2 (2011); 
P. Zhang, and M. W. Wu, arXiv:1108.0283v2 (2011).







\bibitem{Nature07Meyer}
J. C. Meyer, A. K. Geim, M. I. Katsnelson, K. S. Novoselov, T. J. Booth, and S. Roth, 
Nature (London), {\bf 446}, 60 (2007).

\bibitem{PNAS07Stolyarova}
E. Stolyarova, K. T. Rim, S. Ruy, J. Maultzsch, P. Kim, L. E. Brus, T. F. Heinz, M. S. Hybertsen, and G. W. Flynn, 
Proc. Natl. Acad. Sci. U.S.A. {\bf 104}, 9209 (2007).





\bibitem{NanoLett07Ishigami}
M. Ishigami, J. H. Chen, W. G. Cullen, M. S. Fuhrer, and E. D. Williams, 
Nano Lett. {\bf 7}, 1643 (2007). 

\bibitem{PRL09Geringer}
V. Geringer, M. Liebmann, T. Echtermeyer, S. Runte, M. Schmidt, R. R\"{u}ckamp, M. C. Lemme, and 
M. Morgenstern, Phys. Rev. Lett. {\bf 102}, 076102 (2009).





\bibitem{PRL10Cullen}
W. G. Cullen, M. Yamamoto, K. M. Burson, J. H. Chen, C. Jang, L. Li, 
M. S. Fuhrer, and E. D. Williams, Phys. Rev. Lett. {\bf 105}, 215504 (2010). 

\bibitem{PRB11Knox}
K. R. Knox, A. Locatelli, M. B. Yilmaz, D. Cvetko, T. O. Mente\c{s}, M. \'{A}. Ni\~{n}o, P. Kim, A. Morgante, and R. M. Osgood Jr., Phys. Rev. B {\bf 84}, 115401 (2011). 



\bibitem{NatNano09Bao}
W. Bao, F. Miao, Z. Chen, H. Zhang, W. Jang, C. Dames, and C. N. Lau, Nat. Nanotechnol.
{\bf 4}, 562 (2009).

\bibitem{Science11Choi} 
J. S. Choi, J.-S. Kim, I.-S. Byun, D. H. Lee, M. J. Lee, B. H. Park, C. Lee, D. Yoon, H. Cheong, K. H. Lee, Y.-W. Son, J. Y. Park, and M. Salmeron, Science, {\bf 333}, 607 (2011).





\bibitem{PRL09Huertas-Hernando}
D. Huertas-Hernando, F. Guinea, and A. Brataas, Phys. Rev. Lett. 
{\bf 103}, 146801 (2009).

\bibitem{PRB10Zhou}
Y. Zhou and M. W. Wu, Phys. Rev. B {\bf 82}, 085304 (2010). 

\bibitem{PRB11Dugaev}
V. K. Dugaev, E. Ya. Sherman, and J. Barna\'{s}, Phys. Rev. B {\bf 83}, 085306 (2011). 


\bibitem{Nature08Kuemmeth}
F. Kuemmeth, S. Ilani, D. C. Ralph, and P. L. McEuen, Nature (London) {\bf 452}, 448 (2008).



\bibitem{JPSJ00Ando}
T. Ando, J. Phys. Soc. Jpn. {\bf 69}, 1757 (2000).

\bibitem{JPSJ09Izumida}
W. Izumida, K. Sato, and R. Saito, J. Phys. Soc. Jpn. {\bf 78}, 074707 (2009).



\bibitem{PRB09Jeong}
J.-S. Jeong and H.-W. Lee, Phys. Rev. B {\bf 80}, 075409 (2009).




\bibitem{NatPhys11Jespersen}
T. S. Jespersen, K. Grove-Rasmussen, J. Paaske, K. Muraki, T. Fujisawa, J. Nyg{\aa}rd, 
and K. Flensberg, Nat. Phys. {\bf 7}, 348 (2011).



\bibitem{NatMat07Fasolino}
A. Fasolino, J. H. Los, and M. I. Katsnelson, Nature Mater. {\bf 6}, 858 (2007).




\bibitem{flatSG}
It was reported recently that
the corrugation can be suppressed if the SG is exfoliated
onto mica surfaces
[C. H. Lui, L. Lui, K. F. Mak, G. W. Flynn, and T. F. Heinz, Nature (London), {\bf 462}, 339 (2009)]. 






\bibitem{PTTSA08Katsnelson}
M. I. Katsnelson and A. K. Geim, Phil. Trans. R. Soc. A {\bf 366},
195 (2008).




\bibitem{PRB84DiVincenzo}
G. W. Semenoff, Phys. Rev. Lett. {\bf 53}, 2449 (1984);
D. P. DiVincenzo and E. J. Mele, Phys. Rev. B {\bf 29}, 1685 (1984). 

\bibitem{SSC00Serrano}
J. Serrano, M. Cardona, and T. Ruf, Solid State Commun. {\bf 113}, 411 (2000).


\bibitem{PRL05Kane}
C. L. Kane, and E. J. Mele, Phys. Rev. Lett. {\bf 95}, 226801 (2005).


\bibitem{PRL91Tomanek}
D. Tom\'{a}nek and M. A. Schluter, Phys. Rev. Lett. 
{\bf 67}, 2331 (1991).

\bibitem{Book88Struik}
D. J. Struik, {\it Lectures on Classical Differential Geometry}, 2nd ed.
(Dover, New York, 1988).



\bibitem{Book68Schiff}
Leonard I. Schiff, {\it Quantum Mechanics} (McGraw-Hill, New York, 1968). 




\bibitem{PRB09Gmitra}
M. Gmitra, S. Konschuh, C. Ertler, C. Ambrosch-Draxl, and J. Fabian, Phys. Rev. B {\bf 80}, 235431 (2009).

\bibitem{PRB10Konschuh}
S. Konschuh, M. Gmitra, and J. Fabian, Phys. Rev. B {\bf 82}, 245412 (2010).



\bibitem{EPL08Kim}
E.-A. Kim and A. H. Castro Neto, Europhys. Lett. {\bf 84}, 57007 (2008).
 
 
\bibitem{PRB06Min}
H. Min, J. E. Hill, N. A. Sinitsyn, B. R. Sahu, L. Kleinman, and A. H. MacDonald, Phys. Rev. B {\bf 74}, 165310 (2006);
Y. Yao, F. Ye, X.-L. Qi, S.-C. Zhang, and Z. Fang, {\it ibid} {\bf 75}, 041401(R) (2007).
 

\bibitem{Second-order perturbation}
The 
comprehensive
explanation of the degenerate second-order perturbation theory 
is described well in Ref.~\cite{PRB06Huertas-Hernando,PRB09Jeong}. 


\bibitem{PR54Slater}
J. C. Slater and G. F. Koster, Phys. Rev. {\bf 94}, 1498 (1954). 












\bibitem{JPhys02Averkiev}
{\it Optical Orientation}, edited by F. Mayer and B. Zakharchenya (North-Holland, Amsterdam, 1984);
N. S. Averkiev, L. E. Golub, and M. Willander, J. Phys.:Condens. Matter {\bf 14}, 
R271 (2002); 
S. A. Tarasenko, JETP Letters {\bf 84}, 199 (2006).






\bibitem{JPSJ06Ando}
T. Ando, J. Phys. Soc. Jpn. {\bf 75}, 124701 (2006),

\bibitem{PRL06Morozov}
S. V. Morozov, K S. Novoselov, M. I. Katsnelson, F. Schedin, 
L. A. Ponomarenko, D. Jiang, and A. K. Geim, Phys. Rev. Lett. 
{\bf 97}, 016801 (2006).

\bibitem{PRB08Guinea}
F. Guinea, B. Horovitz, and P. Le Doussal, Phys. Rev. B {\bf 77}, 205421 (2008).



\bibitem{electron_hole_symmetry}
We evaluate SRT for hole states of Eq.~(\ref{eq:DiracH[1]}) as well,
and obtain electron-hole symmetric SRT. 




\bibitem{no correlation}
A strain-induced vector potential  
arising from in-plane atomic displacement~\cite{JPSJ06Ando,PRL06Morozov}
is not correlated with the ${\cal H}_{\rm soc}({\bf r})$ that is dependent on the curvature only. 
In addition, the scalar potential induced by the curvature effects~\cite{EPL08Kim}
is quadratic in local mean curvature.






\bibitem{PRB93Palasantzas}
B. B. Mandelbrodt, {\it The Fractal Geometry of Nature
} (Freeman, New York, 1982); 
G. Palasantzas, Phys. Rev. B {\bf 48}, 14472 (1993).


\bibitem{Communications82Fournier}
A. Fournier, D. Fussel, and L. Carpenter, Communications of the ACM, {\bf 25}, 371 (1982).


\bibitem{JPSJ07Ando}
T. Ando, J. Phys. Soc. Jpn. {\bf 75}, 074716 (2006);
K. Nomura and A. H. MacDonald, Phys. Rev. Lett. {\bf 96}, 256602 (2006);
K. Nomura and A. H. MacDonald, {\it ibid} {\bf 98}, 076602 (2007); 
E. H. Hwang, S. Adam, and S. Das Sarma, {\it ibid} {\bf 98}, 186806 (2007);
S. Adam, E. H Hwang, V. Galitski, and S. Das Sarma, Proc. Natl. Acad. Sci. U. S. A.
{\bf 104}, 18392 (2007).









\end{thebibliography}
\end{document}